\documentclass[prl,preprint,groupedaddress]{revtex4}
\usepackage{dcolumn}
\usepackage{graphics}
\usepackage{amsmath}
\usepackage{bm}
\usepackage[usenames]{color}
\begin{document}

\title{Nanopolaritons: Vacuum Rabi splitting with a
single quantum dot in the center of a dimer nanoantenna}

\author{S. Savasta, R. Saja, A. Ridolfo, O. Di Stefano, P. Denti, F. Borghese}
\affiliation{Dipartimento di Fisica della Materia e Ingegneria Elettronica, Universit\`{a} di Messina Salita Sperone 31, I-98166 Messina, Italy}
\date{\today}

\begin{abstract}
{The demonstration of enhanced spontaneous emission of nanoscaled optical emitters near metallic nanoparticles and the recent realization of a nanolaser based on surface plasmon amplification by stimulated emission of radiation (spaser) encourage the search for strong coupling regime at the nanoscale.
Here we propose the concept of nanopolaritons. We demonstrate with accurate scattering calculations that the strong coupling regime of a single quantum emitter (a semiconductor quantum dot) placed in the gap between  two  metallic nanoparticles can be achieved. The largest dimension of the investigated  system  is only 36 nm.
Nanopolaritons will advance our fundamental understanding of surface plasmon enhanced optical interactions and  could be used as ultra-compact elements in quantum-information technology.}
\end{abstract}

\maketitle

Cavity quantum electrodynamics addresses properties of quantum emitters in microcavities and can be divided into a weak and a
strong coupling regime. For weak coupling, the spontaneous emission can be enhanced or reduced compared with its vacuum level \cite{Vahala,Goy,Gabrielse,Bayer}. However, the most striking change of emission properties occurs when the conditions for strong coupling are fulfilled. In this case there is a change from the usual irreversible spontaneous emission to a reversible exchange of energy between the emitter and the cavity mode giving rise to energy splitted half-light half-matter modes
(also known as cavity-polaritons) \cite{Mabuchi, Raimond, Weis, Yoshie, Reit}.  This nonperturbative regime is highly desirable
for many possible applications in quantum information processing or schemes for ultrafast coherent control \cite{Monroe,Duan}.
Applying cavity quantum electrodynamics (QED) techniques to quantum information requires a single-quantum emitter/single-photon coupling that overwhelms any loss or decoherence rate, including atomic spontaneous emission and photon leakage through the mirrors \cite{Monroe}. 
If the strength of the coupling between the quantum emitter and the microcavity overcomes the losses, the two subsystems are able to exchange coherently energy before relaxation even at low excitation intensities. In this case  vacuum Rabi oscillations in the time domain and single-quantum frequency splitting (vacuum Rabi splitting) occurs. 
Realizing these tasks in the solid state is clearly desirable, and coupling semiconductor quantum dots to monolithic optical cavities is a promising route to this end \cite{Ima, Fushman}.
An inherent limitation of the cavity induced strong coupling regime  is that the size of the cavity is at least half wavelength and practically much more than that owing to the presence of mirrors or of a surrounding photonic crystal.
Metallic nanoparticles and metallic nanostructures can beat the diffraction limit and focus electromagnetic waves to spots much smaller than a wavelength. In this way it is possible to increase the local density of the electromagnetic modes as in microcavities but with ultra-compact structures, enabling spontaneous emission control of optical transitions \cite{Farahani,Sandoghar,Kinkhabwala}.
This ability stems from the existence of collective, wave-like motions of free electrons on a metal surface, termed surface plasmons. On the other hand, owing to dissipative losses in metals, surface plasmon (SP) modes display very fast relaxation  times $1/\gamma_{\rm SP} \sim 10 - 100$ fs   which can be very helpful for ultrafast signal processing but lowers their performances as effective resonators, so discouraging the  search for strong coupling between localized SPs and quantum emitters at the nanoscale. 
An outstanding demonstration of this cavity-like behavior of metallic nanoparticles is the recent realization of a nanolaser based on surface plasmon amplification by stimulated emission of radiation (spaser) \cite{Spaser}.
Despite the large losses exhibited by SPs, the Rabi splitting between  extended SP waves and organic excitons was demonstrated in Refs. \cite{Pockrand,Bellessa,Hakala}. In these cases the strongly coupled system involves many quantum emitters and extends over many optical wavelengths. Vacuum Rabi splitting has also been observed in hybrid exciton-plasmonic crystals which consist of gold nanovoids (diameter $d=600$ nm) covered with an organic film \cite{voids}.
Recently the efficient coupling between an individual optical emitter and propagating SPs confined to a conducting nanowire  has been studied both theoretically \cite{Chang} and experimentally \cite{Akimov}. Although this system does not display vacuum Rabi splitting, light escaping from the emitter is efficiently transferred to the nanowire. The potential of this system as  an ultracompact single-photon transistor has been theoretically demonstrated \cite{Chang2}.

Here we investigate the requirements to be satisfied by SP-nanosystems coupled to one ore more quantum emitters in order to display the vacuum Rabi splitting. We present accurate calculations demonstrating that a silver dimer nanoantenna coupled to a single quantum dot displays the  Rabi doublet.
We start considering a metal nanoparticle or nanostructure surrounded by the resonant medium that spatially overlaps
with the SP eigenmode  and whose emission
line at energy $\hbar \omega_0$ spectrally overlaps with the SP eigenmode. In the following we will address two different kinds of active media: i) a dielectric matrix doped with
$N$ quantum emitters,  e.g. dye molecules (Fig.\ 1a); ii) a single semiconductor quantum dot  (Fig.\ 2a). In a simplified picture of coupled harmonic oscillators \cite{Rudin}, where the SP resonance is described as a single mode with a frequency independent decay rate $\gamma_{\rm SP}$ \cite{Bergman}, the resonant ($\omega_{\rm SP} = \omega_0$) interaction of the SP mode with one optically active electronic transition gives rise to modes with energy 
\begin{equation}
\Omega_{\pm} = \omega_0 -i(\gamma_{\rm SP} +\gamma_0)/4  \pm \sqrt{g^2-(\gamma_{\rm SP}-\gamma_0)^2 /16}\, ,
\label{split}
\end{equation}
where $\gamma_{\rm 0}$ is the full width at half maximum (FWHM) of the quantum emitters emission line and $g$ is the active medium-SP coupling rate. It is proportional to the dipole moment associated to the transition and to the SP mode density at the transition energy. If the SP-mode interacts with $N$ quantum emitters, assuming that all of them experience the same SP field intensity, and that the coupling for one emitter is $g_1$, the resulting total coupling increases according to $g = \sqrt{N}g_1$. The vacuum Rabi splitting between the two energies Re$\left(\Omega_+\right)$ and Re$\left(\Omega_-\right)$ appears when $g^2 > (\gamma_{\rm SP}-\gamma_0)^2 /16$. In addition, in order to see an evident splitting or complete oscillations before decay in the time domain the further condition $g > (\gamma_{\rm SP}+\gamma_0) /4$ should hold too.
The intrinsic value of the FWHM $\gamma_0$ of the electronic transitions in many cases of interest is much smaller than $\gamma_{\rm SP} \geq  50$ meV. The criterion for strong coupling can therefore be approximated by $g > \gamma_{\rm SP}/4$. Despite the linewidth of SP modes is very large, this criterion can be satisfied thanks to the substantial increase in the
coupling strength $g$ due to the concentration of photons in nanosized volumes enabled by SPs.
Silver nanoparticles  perform much better than their gold counterparts since the imaginary part of the Ag dielectric function drops to much lower values. Moreover the coating of such particles with a dielectric medium can be exploited to shift  the dipole plasmon resonance at longer wavelengths, where the imaginary part of the dielectric function is lower and where it  is more likely to find resonant quantum emitters. 
\begin{figure}[!] \vspace{0. cm}
\begin{center}
\resizebox{15. cm}{!}{\includegraphics{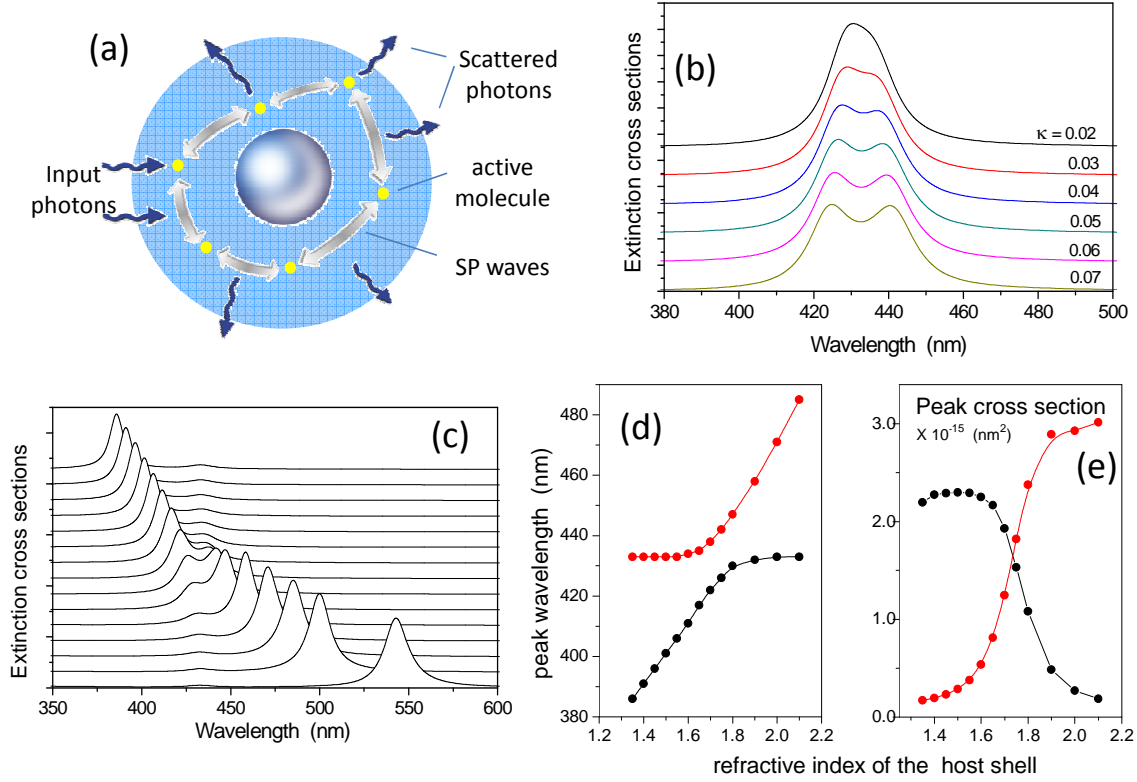}}\caption{ Vacuum Rabi splititng with a silver core interacting with a dielectric shell doped with active organic molecules. {\bf a} Schematic of the strong interaction between surface plasmons
and the active organic molecules in the surrounding shell. The input light excites either the SP waves or the organic molecules.
In each case the excitation is coherently exchanged between them before the excitation is scattered out.{\bf b} Calculated extinction cross sections as function of the wavelength of the input field obtained for different extinction coefficients $\kappa$ of the doped shell. A significant Rabi splitting appears for $\kappa \geq 3 \times 10^{-2}$. {\bf c} Extinction cross sections spectra obtained for different refractive indexes $n_{\rm b}$ of the dielectric host shell. A clear anti-crossing is observed owing to strong coupling between the organic molecules and the localized surface plasmon mode. {\bf d} Dependence of the two Rabi  peak-wavelengths on the refractive index of the host shell $n_{\rm b}$. {\bf e} Dependence of the two Rabi peak extinction cross sections on the refractive index of the host shell $n_{\rm b}$.}
\end{center} \vspace{-0. cm}
\end{figure}
At resonance ($\omega_{\rm SP} = \omega_0$), the interaction with a resonant field-mode  increases  the spontaneous emission rate of the emitter according to the relation \cite{book}
\begin{equation}\label{Gamma1}
\Gamma = \Gamma_0 + 4g^2_1/\gamma_{\rm SP}\, ,
\end{equation}
where $\Gamma_0$ is the spontaneous decay rate in the absence of the metallic nanoparticle.
The spontaneous emission rate $\Gamma$ of a quantum emitter nearby a metallic nanoparticle can also be expressed as  \cite{Farahani,Novotny}
\begin{equation}\label{Gamma2}
\Gamma = \Gamma_0 \rho({\bf r},\omega_0)\, ,
\end{equation}
where  $\rho({\bf r},\omega_0) = D({\bf r},\omega_0)/ D_0(\omega_0)$ is the enhancement of the field mode density at the position and at the transition energy of the emitter with respect to the free space value. From  Eqs\, (\ref{Gamma1}) and (\ref{Gamma2}) one can easily evaluate $g_1$ from the knowledge of  $\Gamma_0$, $\gamma_{\rm SP}$ and $\rho$:
\begin{equation}
g^2_1 = \Gamma_0  \gamma_{\rm SP} (\rho({\bf r},\omega_0) -1)/4\, .
\end{equation}  
For example typical dye molecules have free-space radiative decay times $\tau_{\rm SE} \sim 4$ ns corresponding to 
$\Gamma_0 = \hbar/\tau_{\rm SE} = 2.4\; \mu$eV. A silver nanosphere of radius $a = 7$ nm displays a mode density of the order $\rho(r = 10\, {\rm nm})\, \sim 2000$, where $R$ is the distance from the center of the sphere. The FWHM of the Ag SP dipole resonance coated by a dielectric shell with refractive index $n \sim 1.7$ is about $\gamma_{\rm SP} \sim 60 $ meV. From such data it results $g_1 \sim 2.5$ meV corresponding to weak coupling regime. Nevertheless, it is sufficient  doping a dielectric shell surrounding the nanosphere with $N \geq 50$ molecules in order to achieve strong coupling. 
This analysis also can show that strong coupling regime with a single quantum emitter could  also be achieved, although more demanding. Semiconductor  quantum dots can display radiative decay rates of the order of $\tau = 400$ ps. In this case a mode density of the order of $\rho > 9000$ is needed to achieve strong coupling with a single quantum emitter. This  or even larger values of mode density can be obtained by placing the quantum dot in the gap between couples of closely spaced nanoparticles \cite{ZLi}.
It is interesting to compare the strong coupling condition with the laser (or in the present case spaser) threshold condition at resonance \cite{Spaser}:
$g^2 \Delta > \gamma_{\rm SP} \gamma_0 /4$, where $0<\Delta <1$ is the difference between the inverted population fractions associated to the resonant transition. This condition is much less demanding than the condition for strong coupling when $\gamma_0 \ll \gamma_{\rm SP}$.

The above analysis provides an estimate of the requirements for achieving the vacuum Rabi splitting at the nanoscale but suffers from a number of approximations and oversimplifications: i) broad SP resonances cannot be fully modeled as ideal single mode resonances with constant damping; ii) the interaction with quantum emitters can switch on multipolar contributions which alters the SP density of states; iii) The strong gradients displayed by the localized SP fields as well as polarization effects prevents the possibility for the quantum emitters to experience the same field intensity.
In the following we present detailed scattering calculations for  two different systems: a) nanoparticles with a silver spherical core covered by a  dielectric shell doped with active organic molecules; b) a single quantum dot in between a pair of silver nanospheres.
The optical properties of these  coupled systems can be exactly calculated through the formalism of the multipole expansion of the fields \cite{BorgheseBook}. This formalism based on generalizations of the Mie theory \cite{Bohren} is indeed able to take into account all the multiple scattering processes that occur among the involved scatterers. In particular the optical properties of core/shell spheres are calculated using the extension of the Mie theory to radially non-homogeneous spheres by Wyatt \cite{Wyatt}.
We consider as incident field a monochromatic linearly polarized plane wave.
The scattering cross-section and absorption cross-sections are defined via Poynting's theorem \cite{Bohren}. The scattering cross-section $\sigma_{\rm scat}$ is defined as the total integrated power contained in the scattered field normalized by the irradiance of the incident field and the absorption cross-section $\sigma_{\rm abs}$ is defined by the net flux through a surface surrounding the scattering system normalized by the incident field irradiance, and is thus a measure of how much energy is absorbed by the system. In  the following we calculate extinction cross sections $\sigma_{\rm ext} = \sigma_{\rm scat}+ \sigma_{\rm abs}$ as a function of the incident-field wavelength, being extinction spectroscopy a widely adopted technique for the characterization of nano- and micro-particles.
We employ a silver core of radius $r_{\rm Ag} = 7$ nm, whose frequency-dependent dielectric permittivity is taken from Ref. \cite{JC}, surrounded by a doped dielectric shell giving rise to a whole structure of radius $r = 22$ nm.
The dielectric shell is described in its simplest way as a medium with a single resonance at the energy $E_0$ by the following permittivity,
\begin{equation}\label{per}
	\epsilon = \epsilon_{\rm b} + \frac{A}{E_0^2- E^2 - 2i E \gamma_0}
\end{equation}
where $\epsilon_{\rm b}$ is the background dielectric constant mainly due to the host matrix medium,  $E_0$ is the dispersionless exciton energy, $\gamma_0$ is the total homogeneous plus inhomogeneous broadening of the excitonic resonance, and the constant A is proportional to the oscillator strength of the transition. It depends on the dipole moment of the single active  molecule and on the density of doping molecules. We use parameters corresponding to the organic molecules tetra-(2,6-t-butyl)phenol-porphyrin zinc (4TBPPZn) exploited for the first demonstration of the strong exciton-photon coupling in an organic semiconductor microcavity: $E_0 = 2.88$ eV corresponding to a wavelength $\lambda_0 \sim 430$ nm and  $\gamma_0 = 57$ meV.  The concentration used in Ref\cite{Lidzey} corresponds to $A \sim 7.8 \times 10^{-2}$ (eV)$^2$, giving a peak extinction coefficient $\kappa = 7 \times 10^{-3}$. 
Fig.\ 1b displays extinction cross section spectra as a function of the wavelength of the input field for different extinction coefficients $\kappa$ displayed in the figure. Increasing the extinction coefficient (e.g by increasing the molecules density), the spectra evolve from a single peak at the resonance wavelength of the SP-mode to a doublet characteristic of the strong coupling regime.
We adopted a matrix with background dielectric constant  $\epsilon_{\rm b} = n^2_{\rm b} = 3.01$ able to red-shift the SP dipole mode resonance of the silver nanoparticle at resonance with $E_0$. This shift also determines an useful reduction of the SP mode linewidth thanks to the reduction of $\kappa_{\rm Ag}$ at larger wavelengths.
In order to investigate the coupling between the SP mode and the active molecules, the energies of the two subsystems  must to be tuned through each other. For tuning through resonance, different methods could be employed. Here we simply study structures with different values of $n_{\rm b}$ corresponding to dielectric  matrices with different refractive indexes. Figure 1c shows spectra calculated with different background refractive indexes ranging from 1.35 to 2.1.
Over the entire range of refractive indexes the energies of the two contributions to the spectrum are well separated and avoid
crossing each other. This anticrossing behavior which is more evident  in Fig.\ 1d is characteristic of true strong coupling, the regime of reversible exchange of energy back and forth between the quantum emitters and the localized SP-mode that is, vacuum Rabi oscillations.
It is interesting to observe that at large detuning (e.g. for $n_{\rm b} = 2.1$) the extinction cross section of excitons at $\lambda_0 = 430$ nm  is smaller by more than one order of magnitude than that of the SP-mode. This is consequence of the very strong absorption and scattering cross sections of SP resonances as compared to that of the shell, although doped with resonant molecules. Only at resonance thanks to the strong coupling the two peaks display the same intensity both corresponding to mixed SP-exciton modes as shown in Fig.\ 1(e). Calculations here presented focus on the scattering process, however we expect as well striking modifications (beyond perturbative nanoantennas currently investigated \cite{Sandoghar}) of the active molecules fluorescence induced by the strong coupling  effect. In particular we expect that the strong coupling will transfer to the light emitted by the active molecules (or dots) the broad spectral and giant scattering cross sections of metallic nanoparticles.
In addition  these systems can allow the realization of the  spaser in the strong coupling regime \cite{Cris}.
\begin{figure}[!] \vspace{-0. cm}
\begin{center}
\resizebox{15. cm}{!}{\includegraphics{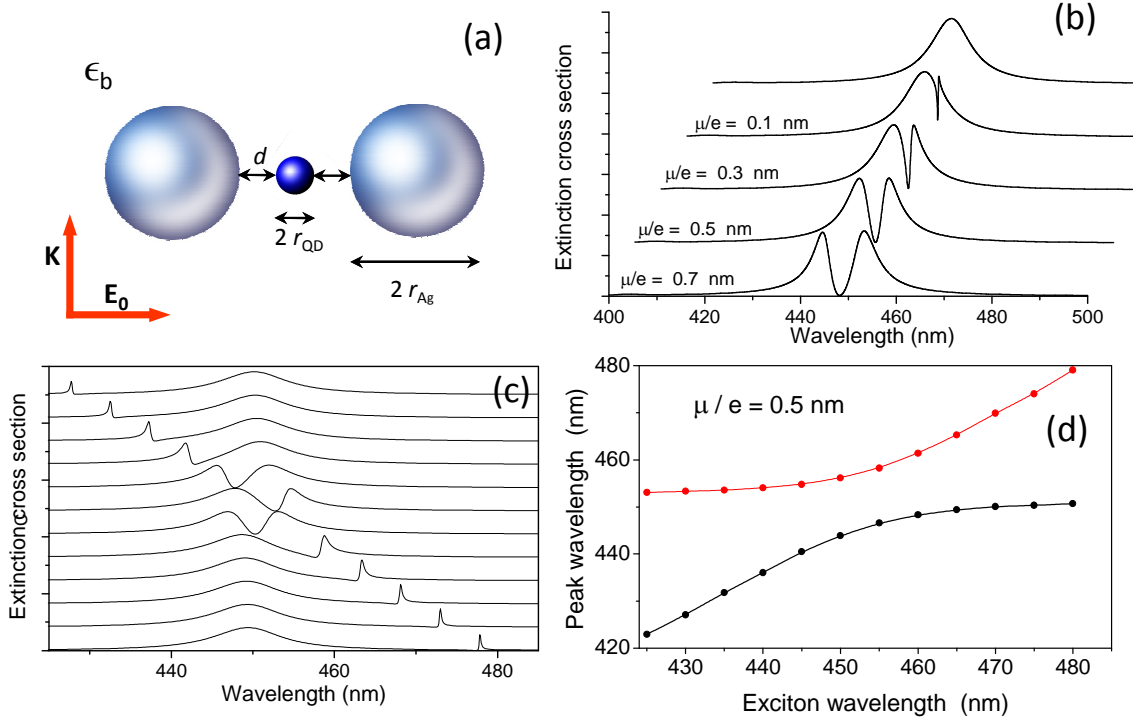}}\caption{ Vacuum Rabi splitting with a single quantum dot in the center of a dimer nanoantenna. {\bf a} Sketch  of the system and of the excitation.
{\bf b} Calculated extinction cross sections as function of the wavelength of the input field obtained for different dipole moments of the quantum dot. {\bf c} Extinction cross sections spectra obtained for different resonant energies $E_0$ of the quantum dot exciton ($\mu/e = 0.5$ nm). A clear anti-crossing is observed owing to strong coupling between the dot and the localized bonding surface plasmon dimer-mode. {\bf d} Dependence of the two Rabi-peaks (extinction cross sections) wavelengths on the exciton transition wavelength $\lambda_0 = h\, c/E_0$ ($\mu/e = 0.5$ nm).}
\end{center} \vspace{-0. cm}
\end{figure}

For a number of quantum information tasks involving quantum operations on single qubits the vacuum Rabi splitting with a single quantum emitter is highly desirable. Semiconductor quantum dots, have dipole moments 50-100 times larger than those of atoms and molecules, while at the same time behaving very close to ideal two-level quantum emitters \cite{Gammon}. Nevertheless our analysis based on Eq.\ (4) showed that placing a single quantum dot a few nanometers close to a metallic nanoparticle is not sufficient to produce  vacuum Rabi splitting. However it is well known that the electromagnetic field in the gap region of a pair of strongly coupled nanoparticles can be drastically amplified, resulting in an extraordinary enhancement factor large enough for single-molecule detection by surface enhanced Raman scattering (SERS)
\cite{ZLi,Nie, Hao, Li}. We exploit this so-called hot spot phenomenon in order to demonstrate that the vacuum Rabi splitting with a single quantum emitter within a subwavelength nanosystem can be achieved. We employ a pair of silver spheres of radius $r_{\rm Ag} = 7$ nm separated by a gap $d = 8$ nm embedded in a dielectric medium with permettivity $\epsilon_{\rm r} = 3$. Individual nanoparticle plasmons hybridize to give two new splitted modes : a bonding and an antibonding combination \cite{Nordlander}. The net dipole moment of the antibonding configuration is zero, this mode is not easily excited by light (dark mode). In contrast, the bonding configuration corresponds
to two dipole moments moving in phase. It is easily excited by input light and produces an extraordinary enhancement of the field mode density  in the gap between the particles. We consider a spherical quantum dot with radius $r_{rm QD} = 2$ nm, whose lowest energy exciton is resonant with the dimer bonding mode. Because of its symmetry, a spherical quantum dot has three
bright excitons with optical dipoles parallel to the three direction x, y, and z respectively. The resulting frequency dependent permettivity is described by Eq.\ (\ref{per}) with $\epsilon_{\rm b} = 3$ and  $A = \mu^2 \hbar / (\epsilon_0 V)$, where $\mu$ is the dipole moment, and $V$ is the dot volume. Fig.\ 2a display a sketch of the system and of the input field polarized along the trimer axis in order to provide the largest field enhancement at the dot position. Fig.\ 2b shows the  extinction cross section spectra calculated for different dipole moments $\mu =  e r_0$, being $e$ the electron charge.
For $r_0 = 0.1$ nm a narrow hole in the spectrum occurs which could be confused with the appearance of a small vacuum Rabi splitting. Actually this hole appears when the exciton linewidth is smaller than the exciton-field coupling strength (or viceversa) which in turn is smaller than the linewidth of the SP mode. This effect can be understood in terms of interference between the continuous SP field and the narrow excitonic resonance \cite{Rice}  giving rise to an anti-resonance or to a Fano-like effect \cite{Bryant}. This interesting intermediate regime of SP-quantum emitter coupling has been  studied theoretically \cite{Chang2, Bryant} and also realized experimentally (for several organic emitters) covering a gold nanoshell with an ultrathin layer of J-aggregate complexes \cite{Plexcitons}.
Only for higher dipole moments $r_0 = 0.3$ a true splitting can be observed. Fig.\ 2c displays the extinction spectra for $r_0 = 0.5$ nm obtained changing the energy of the quantum dot exciton. At large detuning the peak arising from the quantum dot is significantly narrower from that originating from the SP bounding mode. Lowering the detuning increases the linewidth of the exciton-like peak while at the same time lowers that of the SP-like peak as a consequence  of the strong coupling between the modes. Fig.\ 2d  show the Dependence of the two Rabi-peaks (extinction cross sections) wavelengths on the exciton transition wavelength $\lambda_0 = h\, c/E_0$ ($\mu/e = 0.5$ nm). The anticrossing behaviour certifying true strong coupling is evident.
There are several types of quantum dots which efficiently emit light at the wavelengths here adressed, as e.g ZnS and CdS nanocrystals.
In addition the use of different nanostructures as nanoshells and/or embedding dielectric media offer a great potential for tunability.
The anticrossing behavior is clearly evident in Fig. 2d. The dipole moments here employed are typical for small semiconductor quantum dots. Quantum dots with dipole moments  $\mu/e \sim 1.5$ nm have been studied experimentally \cite{Gammon}.

We have shown that the vacuum Rabi splitting between an individual quantum emitter and the light field confined at metallic nanoparticles can be achieved in systems a few tens of nanometers wide. 
These ultracompact quantum systems beating the diffraction limit are in experimental reach. Thanks to the well known quantum optical nonlinear properties of single quantum dots, novel quantum plasmonic devices based on nanopolaritons can allow the implementation of the very reach physics and applications of cavity QED  pushing down dimensions of more than one order of magnitude. The investigated systems are also expected to significantly modify the concept of optical nanoantennas with metallic nanostructures \cite{Sandoghar, Novotny}. Along this road it will be possible to implement scalable photonic (plasmonic) quantum computation \cite{Monroe,Duan} without renouncing to the  nanometric size of the classical logic gates of the present most compact electronic technology.

\end{document}